# Selective chiral symmetry breaking: when "Left" and "Right" cannot coexist and "Left" is the best option.

Cristobal Viedma

*Department of Crystallography and Mineralogy, Faculty of Geology, Complutense University, Madrid, Spain. E-mail: viedma@geo.ucm.es*

**Chiral symmetry breaking occurs when a physical or chemical process, with no preference for the production of one or other enantiomer, spontaneously generates a large excess of one of the two enantiomers: (L), left-handed or (D), right handed. Inorganic processes involving chiral products commonly yield a racemic mixture of both. However life on Earth uses only one type of amino acids (L) and one type of natural sugars (D). The origin of this selective chirality has remained a fundamental enigma in the origin of life since the time of Pasteur, some 140 years ago. Sodium bromate ($NaBrO_3$) and sodium chlorate ($NaClO_3$) when crystallize from an unstirred solution generates statistically equal numbers of left-handed (L) and right handed (D) chiral crystals. But when these two populations of crystals undergo a dissolution-crystallization phenomenon, they cannot coexist: one of them disappears in an irreversible competition process that nurtures the other one. From the viewpoint of energy, these two enantiomers can exist with an equal probability, thus the result of this competition in different systems would be populations of crystals either (L) or (D). But contradicting this theoretical prediction the handedness of the chiral crystals that win the competition and remain in solution in the different systems is almost always the same (99.2 percent for $NaBrO_3$). We suggest that these results are the consequence of Parity-violating energy difference (PVED) between enantiomers and reinforce the idea of a key role of PVDE theory in the origin of biomolecular chirality on Earth.**

**Introduction.**

With a few exceptions the symmetry breaking produced by different natural mechanisms have proved giving small enantiomeric excess (EE) (1), ranging from the 20% found experimentally for asymmetric photolysis, to the $10^{-17}$ alleged theoretically for parity violating energy difference between enantiomers. This means that to reach total chiral purity, mechanisms to enhance any initial imbalance in chirality are absolutely essential (2). In 1953 Frank (3) suggested that a form of autocatalysis in which each enantiomer catalyses its own production, while suppressing that of its mirror image, might have nonlinear dynamics leading to the amplification of small initial fluctuations in the concentrations of the enantiomers. Many theoretical models are proposed afterwards, but they are often criticized as lacking any experimental support (4).

Chiral symmetry breaking, however, is found not only in biological and chemical systems, but also in other systems such as in crystallization. Here we report how so simple a process as dissolution-crystallization generates both autocatalysis and competition and thus produce total and, surprisingly, selective chiral symmetry breaking.

**Sodium chlorate and bromate crystallization**.

The achiral molecules of $NaClO_3$ crystallize as two enantiomeric chiral crystals (5) in the cubic space group $P2_13$. Hence sodium chlorate is achiral before crystallization, as it exists in solution as more or less dissociated ions or clusters without a fixed chirality, but forms a chiral crystal.

When the sodium chlorate crystal grows from a solution while the solution is not stirred, a racemic mixture of D and L crystals emerges. When the solution was stirred all of the crystals had the same random chirality, either levo or dextro (6).

The habitual explanation for this indiscriminate chiral symmetry breaking is that secondary nucleation by which a seed crystal or randomly generated single chiral crystal or 'mother crystal' triggers the production of a large number of secondary crystals at a fast rate if the solution is stirred that are enantiomerically identical to itself. The result of this crystallization process is the generation of crystals with the same handedness in a particular solution. Obviously the handedness of crystals in different solutions is random (6).

But recently we have described a new symmetry breaking process: We show experimental data indicating that complete symmetry breaking and chiral purity can be achieved from an initial system where both enantiomers are present since the beginning; This is an experimental case in which one observes the complete elimination of a chiral population of crystals of a hand in favour of the other one (7).

In that work firstly, we show with laboratory experiments how an isothermal saturated solution (8) of $NaClO_3$ with a large population of D- and L-crystals, moves into complete chiral purity : (i) any small initial crystal enantiomeric excess (CEE) eventually gives rise to total crystal purity disappearing the less abundant enantiomer (100% CEE); (ii) "symmetric" proportion of both enantiomeric crystals (a fifty-fifty mix of chiral crystals) gives rise to total symmetry breaking and crystal purity disappearing randomly one of the two enantiomers.

We stated (7) that in our systems this process becomes possible by the combination of: (i) nonlinear autocatalytic dynamic of secondary nucleation due to the combined abrasion-grinding (glass balls) and stirring in our experiments; and (ii) the recycling of crystallites (Fig.1) when they reach the achiral molecular level in a competitive dissolution-crystallization process. Thus complete chiral purity can be achieved. Up to now there are another two theoretical studies devoted to explain our experiments (9), (10).

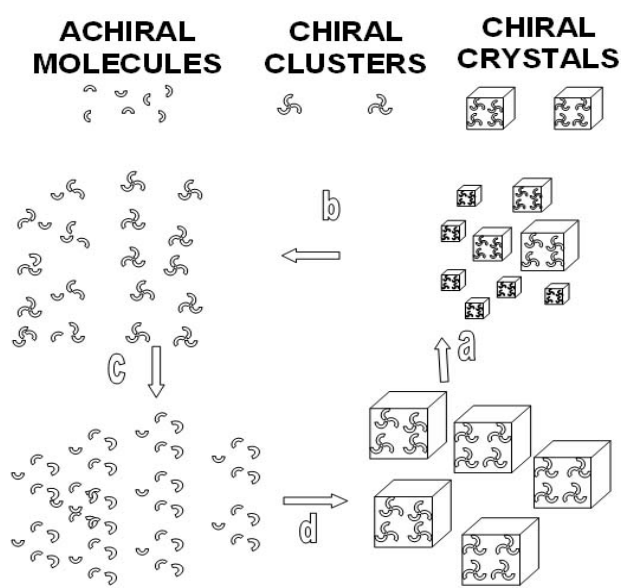

FIG 1. *Recycling process. The abrasion-grinding process generates micro-crystals that easily dissolve (a). The final stage of any crystallite or chiral cluster is the achiral molecular level (b) (c). These molecules feed other crystals independently of its chirality (d).*

NaBrO$_3$ is isomorphous with NaClO$_3$ crystallizing in the enantiomorphic point group 23. Beurskens-Kerssen et al reported (11) that NaClO$_3$ and NaBrO$_3$ of the same chirality have opposite senses of optical rotation (like many aminoacids). We repeat the experiments described in (7) for NaClO$_3$ with NaBrO$_3$. We reach similar results: After a few hours of intense dissolution-crystallization, solutions with initial 5% L-CEE show 100% L-CEE, and solutions with initial 5% D-CEE show 100% D-CEE (Fig.2). 'Symmetric' mixtures show chiral purity and the handedness is L or D randomly.

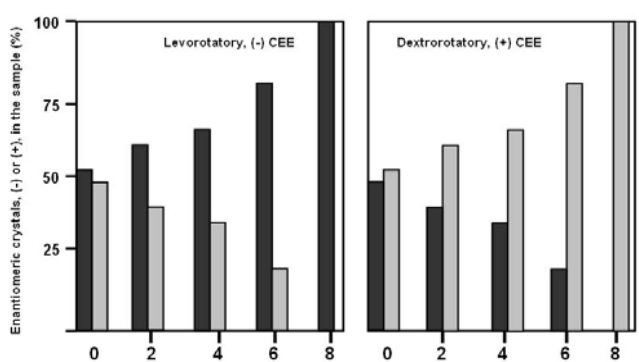

FIG 2. *Solution with initial 5% L-CEE show 100% L-CEE, and solution with initial 5% D-CEE show 100% D-CEE after a few hours.*

When we prepared samples with "symmetric" mixtures of chiral crystals of both NaBrO$_3$ and NaClO$_3$, the result is total symmetry breaking and the handedness is random. However the initial populations of every sample with 'symmetric' mixtures of chiral crystals were always obtained mixing L and D crystals from two different solutions (7). We soon noticed that in this case "any small difference between L and D-crystals induces the preferred production of one of them, for example small differences in the quality of the crystals bias the progressive enantiomeric amplification of a certain handedness" (7). Thus any small difference in the particular history of every solution has a direct effect on characteristics of crystals generated from it and consequently it can bias the results of the experiments when we mix crystals from different solutions. To obviate this inconvenience we design a new experiment in which we obtain both populations of L and D-crystals freely in the same system, it is to say, both populations of chiral crystals generated spontaneously in the same solution under identical circumstances.

**Experimental and results.**

We prepared 10 different solutions of NaClO$_3$ and NaBrO$_3$ by dissolving 8 g of NaClO$_3$ (from Sigma and from Panreac in parallel experiments) in 10 mL of water (bi-distilled water and deionized water (MilliQ-system) in parallel experiments). In the case of NaBrO$_3$ (from Sigma and from Panreac in parallel experiments) we dissolve 4 g in 12 mL of water (bi-distilled water and deionized water (MilliQ-system) in parallel experiments). The resulting solutions was constantly stirred and heated to 100ºC to ensure complete dissolution of the solute and then cooled to 40ºC. These solutions were transferred to Petri dishes of 10 cm of diameter. In order to obtain a statistically relevant numbers of L and D crystals with the same size, these Petri dishes were placed (80 cm) under a continuous flow of air generated by a domestic ventilator of 60 w for at least 24 hours. The initial concentration was such that cooling alone did not produce any crystals, evaporation of the solution was necessary. After this evaporation process, every Petri dish shows an apparently "symmetric" population or mixture of both L- and D-crystals (between 400-600 isometric micro-crystals). The handedness of the chiral crystals was determined by their optical activity using a petrographic microscope (6).

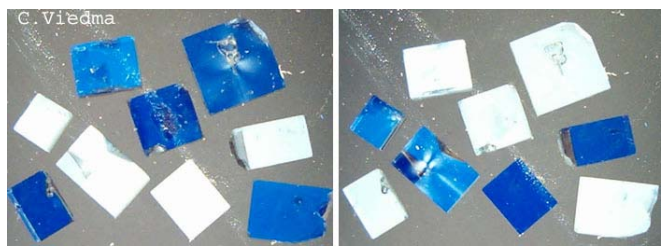

FIG 1 *Light passing through the crystals exhibits optical activity. Left: L crystals show blue colour and D crystals white when we rotate the polarizer a few degrees clockwise. Right L crystals show white colour and D crystals blue when we rotate the polarizer a few degrees counter clockwise.*

Following the experimental protocol described in (7), crystals of every Petri dish was recollected and ground to a fine powder using an agate pestle (almost all systems develop efflorescences on the wall of Petri dishes that are not considered in the evaluation). Then this fine powder was placed in 50 mL round-bottom flasks with 8 g of small glass balls (3-5 mm of diameter). We add 4 mL of water ($NaClO_3$) and 5 mL of water ($NaBrO_3$) that dissolves partially the crystals at the same time that solution, balls and crystals are stirred by a magnetic bar at 800-900 rpm. These experiments were repeated 20 times for $NaClO_3$ and 26 times for $NaBrO_3$ (different experiments from a previous version of this paper). After 24-36 hours total symmetry breaking and complete chiral crystal purity is achieved (Fig. 3).

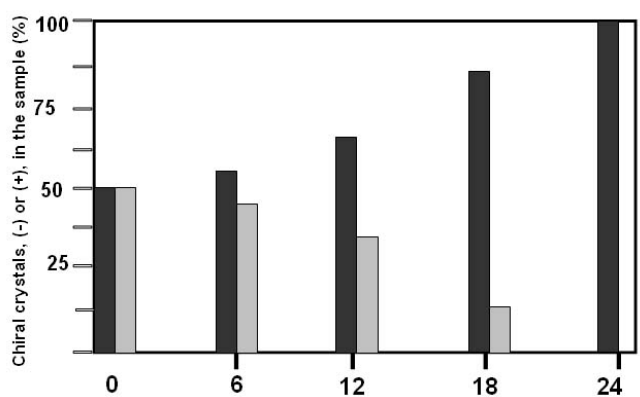

FIG 3. *Initial "symmetric" mixtures of D and L-crystals, generated in the same solution, show total symmetry breaking and chiral purity after 24-36 hours. The handedness of the enantiomer that remains in solution in the different systems is predominantly the same (79.5% for $NaClO_3$ and 99.2% for $NaBrO_3$).*

From 200 experiments with $NaClO_3$, 160 systems show L-crystals and 40 D-crystals (79.5% L-crystals). From 260 experiments with $NaBrO_3$ 258 systems show L-crystals and 2 systems D-crystals (99.2% L-crystals). Although the assignment of chirality is arbitrary we now that both $NaClO_3$ and $NaBrO_3$ show the same selective chiral symmetry breaking because the same structural handedness of chiral crystals has opposite senses of optical rotation. These processes are examples of total and spontaneous symmetry breaking in every solution with chiral selectivity between the different solutions (greater than 99% in one case).

**Discussion.**

We show experimental data indicating that complete homochirality and chiral purity can be achieved from an initial system where both enantiomers of crystals are present in a "symmetric" proportion: in the case of $NaBrO_3$ is the second experimental case in which one observes the complete elimination of a chiral population of crystals of a hand in favour of the other one. The explanation was stated in the first experimental case with $NaClO_3$ (7), thus we are dealing with a general phenomenon for similar compounds that are chiral as crystals but achiral at molecular level. But now, with this new experiment in which we obtain both populations of L and D-crystals in the same solution under identical circumstances, we reached total but selective symmetry breaking with almost always the same handedness of the chiral crystals that win the competition and remain in solution in the different systems. In absence of any chiral physical force, we suggest that only parity violation can account for such a selective chiral symmetry breaking.

The discovery that parity is not conserved by the weak interactions supported a minority chemical tradition deriving from Pasteur, which maintained that there is an intrinsic dissymmetric force inherent in the physical world. Thus PVED can manifest itself in a number of ways, in particular, as an imbalance in the concentrations of enantiomeric species in classically racemic equilibrium, and also as an inequivalence in the rate constant of enantiomeric reactions (12).

The essential difficulty is the minuteness of the intrinsic PVED for small chiral molecules ($10^{-17}$). However in polymerization or crystallization, the influence of this small ratio may increase in proportion for the number of monomers in the polymer or the number of molecules in the crystal according to Yamagata's linear model (13): Giant chiral molecules such as polymers, and especially crystal lattices, may have much larger PVED, and may exhibit the resulting dissymmetries at a detectable level even without the intervention of amplification mechanisms (14). It is to say, in crystallization processes the overall PVED has a linear relationship with the number of unit cells. For the large number of unit cells possible in crystal lattices, even very small non-zero PVED per unit can be enhance to give a detectable difference in the quantities or in the rate constant of left- and right-handed forms of a chiral crystal (14).

The essence of our nonlinear autocatalytic-recycling process is the competence between millions of micro-crystals in a continuous dissolution-crystallization phenomenon with the result of total symmetry breaking and complete chiral purity.

Thus, following the former theoretical predictions, we are dealing with the perfect scenario in which any small difference between enantiomers due to PVED can be checked.

We suggest that the selective chiral symmetry breaking that shows the results of our experiments is consequence of Parity-violating energy difference between enantiomers. This reinforces the idea of a key role of PVDE theory in the origin of biomolecular chirality on Earth.

1. Bonner, W. A. The Origin and Amplification of Biomolecular Chirality, Origin of Life and Evolution of the Biosphere **21**, 59–111. (1991)
2. Schmidt, J. G., Nielsen, P. E. & Orgel, L. E. Enantiomeric cross-inhibition in the synthesis of oligonucleotides on a nonchiral template. *J. Am. Chem. Soc.* **119**, 1494-1495 (1997).
3. F. C. Frank, Biochem. Biophys. Acta 11, 459 (1953).
4. W.A. Bonner, Topics Stereochem. 18, 1 (1988).
5. S. C. Abrahams and J. L. Bernstein, Acta Cryst. B 33, 3601 (1977).
6. Kondepudi, D. K., Kaufman, R. J. & Singh, N., Chiral Symmetry Breaking in Sodium Chlorate Crystallization. Science 250, 975–976. (1990)
7. Viedma C., Chiral symmetry breaking during crystallization: Complete chiral purity induced by nonlinear autocatalysis and recycling. Phys. Rev. Lett. 94, 065504 (2005). (arXiv.Condensed-matter/0407479)
8. Chen W. C., Liu D.D., Ma W.Y.,. Xile A. Y., Fang, The determination of solute distribution during growth and dissolution of $NaClO_3$. J. Cryst.. Grow. 236, 413-419.
9. Uwaha M., A Model for Complete Chiral Crystallization J. Phys. Jpn. 73, 10, 2601 (2004)

10. Saito Y., Hyuga H. Chirality selection in crystallization. J. Phys. Soc. Jpn. 74 (2005) 535-537.
11. Abrahams S.C., Glass A. M., Nassau K. Crystal chirality and optical rotation sense in isomorphous $NaClO_3$ and $NaBrO_3$. Solid State Communication 24, 515-516 (1977).
12. Tranter G. E. The paritiy-violating energy difference between enantiomeric reactions. Chemical physics letters 115, 3, 286-290 (1985).
13. Yamagata Y J. A hypothesis for the asymmetric appearance of biomolecules on earth. J Theor Biol;11(3), 495–498 (1966)
14. Tranter G. E. Paritiy-violating energy differences of chiral minerals and the origin of biomolecular homochirality. Nature, 318,172-173, (1985).